\documentclass[a4paper]{jpconf}
\usepackage{graphicx}
\usepackage{type1cm}
\usepackage{fancyhdr}

\begin{document}
\title{
\vspace*{-3.67cm}\\
{\normalsize\normalfont
DESY 11-217\hfill\mbox{}\\
FTUAM-11-63\hfill\mbox{}\\
IFT-UAM/CSIC-11-91\hfill\mbox{}\\
November 2011\hfill\mbox{}\\}
\vspace{1cm}
Indirect searches for gravitino dark matter}

\author{Michael Grefe}

\address{Deutsches Elektronen-Synchrotron DESY, Notkestra\ss e 85, D-22607 Hamburg, Germany \\[1mm]
         Departamento de F\'isica Te\'orica and Instituto de F\'isica Te\'orica UAM/CSIC, \\
         Universidad Aut\'onoma de Madrid, Cantoblanco, E-28049 Madrid, Spain}

\ead{michael.grefe@uam.es}

\begin{abstract}
The gravitino in models with a small violation of R-parity is a well-motivated decaying dark matter candidate that leads to a cosmological scenario that is consistent with big bang nucleosynthesis and thermal leptogenesis. The gravitino lifetime is cosmologically long-lived since its decays are suppressed by the Planck-scale as well as the small R-parity violating parameter. We discuss the signals in different cosmic-ray species coming from the decay of gravitino dark matter, namely gamma rays, positrons, antiprotons, antideuterons and neutrinos. Comparison to cosmic-ray data can be used to constrain the parameters of the model.
\end{abstract}

\section{Introduction}

The existence of the gravitino in the particle spectrum of locally supersymmetric theories leads to problems in the standard cosmological scenario. One manifestation of this so-called cosmological gravitino problem appears in scenarios where the baryon asymmetry is generated via thermal leptogenesis. The late decays of the abundantly produced gravitino are in conflict with the successful predictions of big bang nucleosynthesis. A similar problem even persists in scenarios where the gravitino is the lightest superpartner and thus a candidate for the dark matter. In this case late decays of the next-to-lightest superpartner can spoil the predictions of big bang nucleosynthesis. Among several other possibilities a very attractive solution to this problem are theories with a slight violation of R-parity~\cite{hep-ph/0702184}.
\thispagestyle{fancy}

\section{Gravitino dark matter with bilinear R-parity violation}

Theories with bilinear R-parity violation only violate lepton number and are thus not affected by proton stability constraints. The size of the R-parity violating couplings are, however, constrained by cosmological considerations: They must be large enough to allow for the decay of the next-to-lightest superpartner before the time of big bang nucleosynthesis and they must be small enough that the baryon asymmetry of the universe is not washed out by these additional interactions. Interestingly, the allowed window for the couplings leads to a gravitino lifetime exceeding the age of the universe by several orders of magnitude, thus making the unstable gravitino a viable dark matter candidate~\cite{arXiv:hep-ph/0005214}.

An intriguing feature of this scenario is that it exhibits a rich phenomenology although the gravitino usually is thought to be one of the most elusive particles. For instance, the next-to-lightest superpartner decays mainly via the small R-parity violating couplings and is thus long-lived on collider scales~\cite{arXiv:1007.5007}. In addition, the decay of gravitino dark matter in the galactic halo can lead to observable signals in the spectra of cosmic rays as we will discuss in the following.

The gravitino can decay into several two-body final states: $\psi_{3/2}\rightarrow\gamma\nu, Z\nu, h\nu$ and $W\ell$. We calculate analytically the branching ratios for these decay channels, while the spectra of stable final state particles -- \textit{i.e.} photons, electrons, protons, deuterons and neutrinos -- are obtained by a PYTHIA simulation of the gravitino decay~\cite{DESY-THESIS-2011-039}. The purpose of this conference contribution is to briefly report on the phenomenological results obtained in~\cite{DESY-THESIS-2011-039}.
\vspace{-3mm}

\section{Indirect detection of gravitino dark matter}

\begin{figure}[b]
  \includegraphics[width=0.49\textwidth]{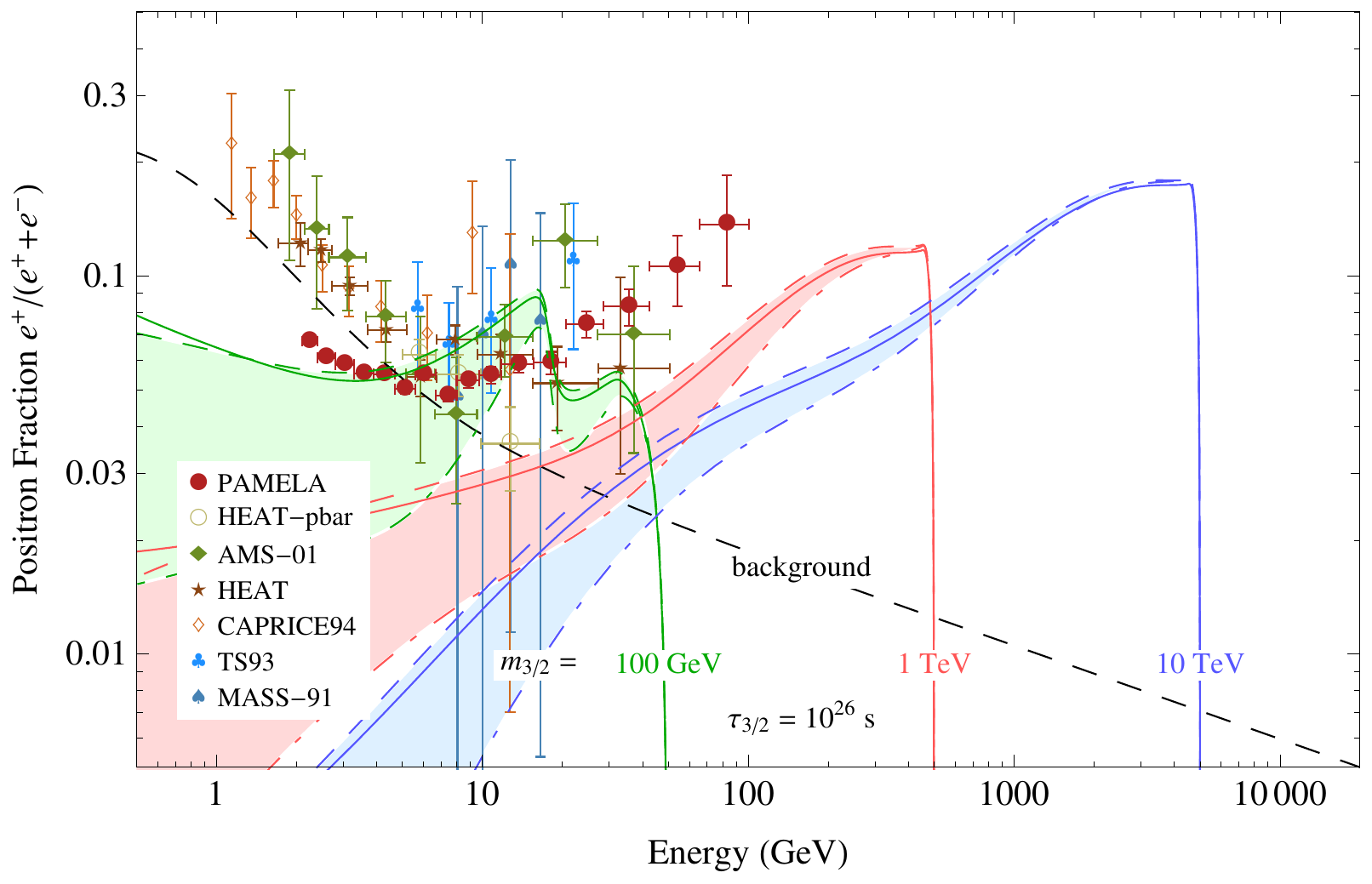}
  \hfill
  \includegraphics[width=0.49\textwidth]{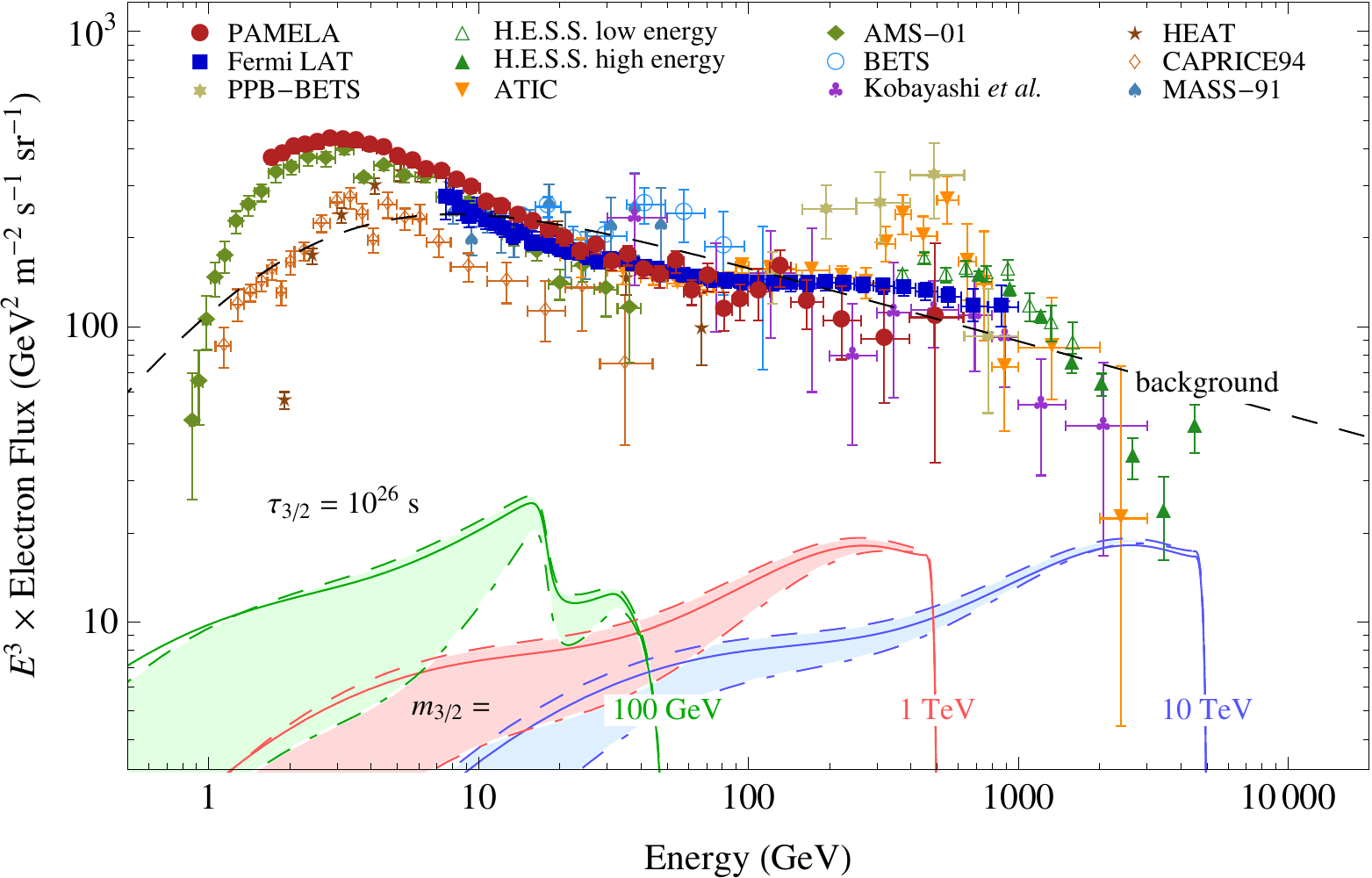}
  \caption{\label{positronelectron}Positron fraction (\textit{left}) and electron spectrum (\textit{right}) expected from gravitino dark matter decays compared to data and the expected astrophysical background. Shaded bands show the uncertainty due to charged cosmic-ray propagation through the galaxy.}
\end{figure}

\begin{figure}[b]
  \includegraphics[width=0.49\textwidth]{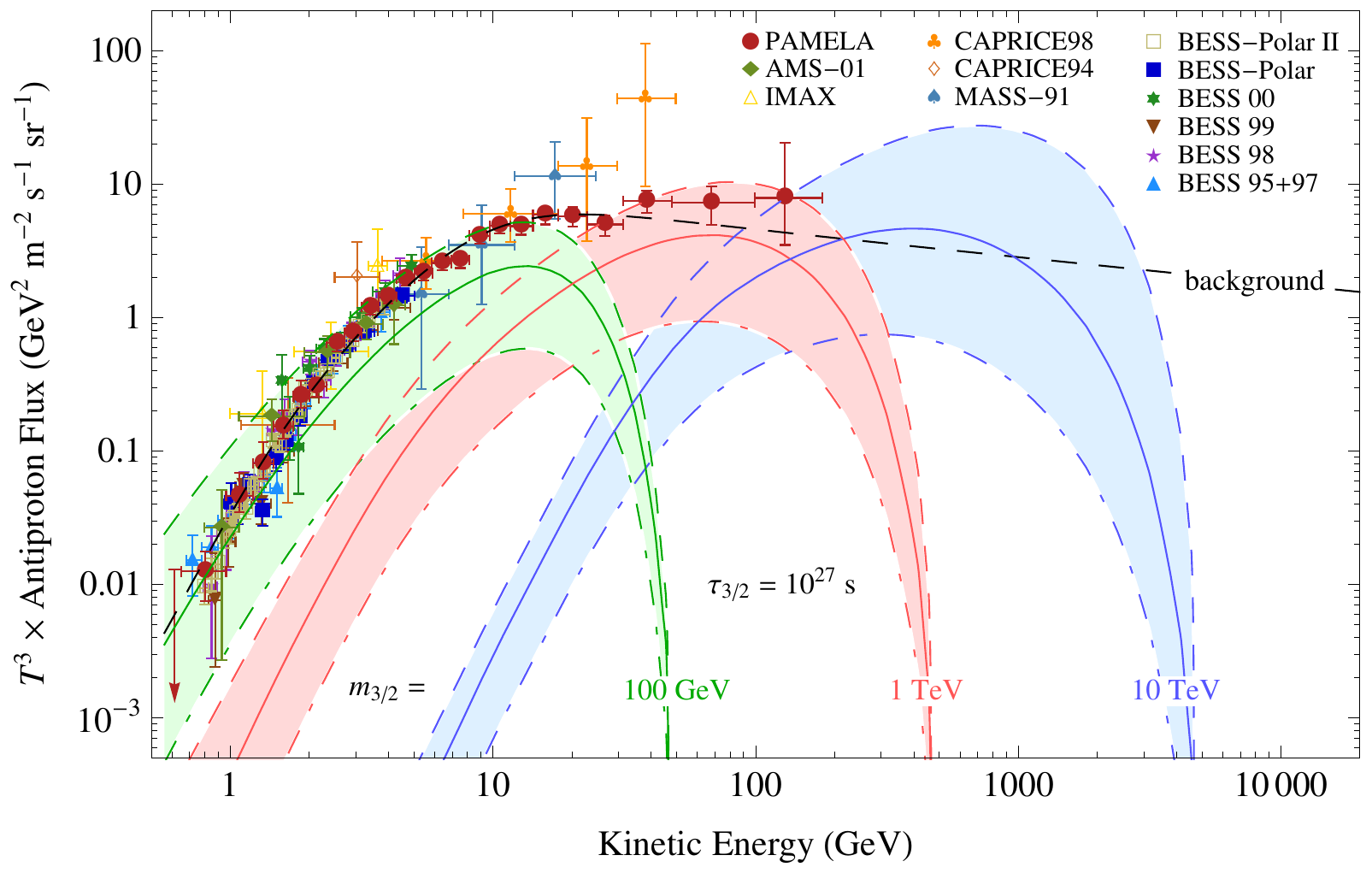}
  \hfill
  \includegraphics[width=0.49\textwidth]{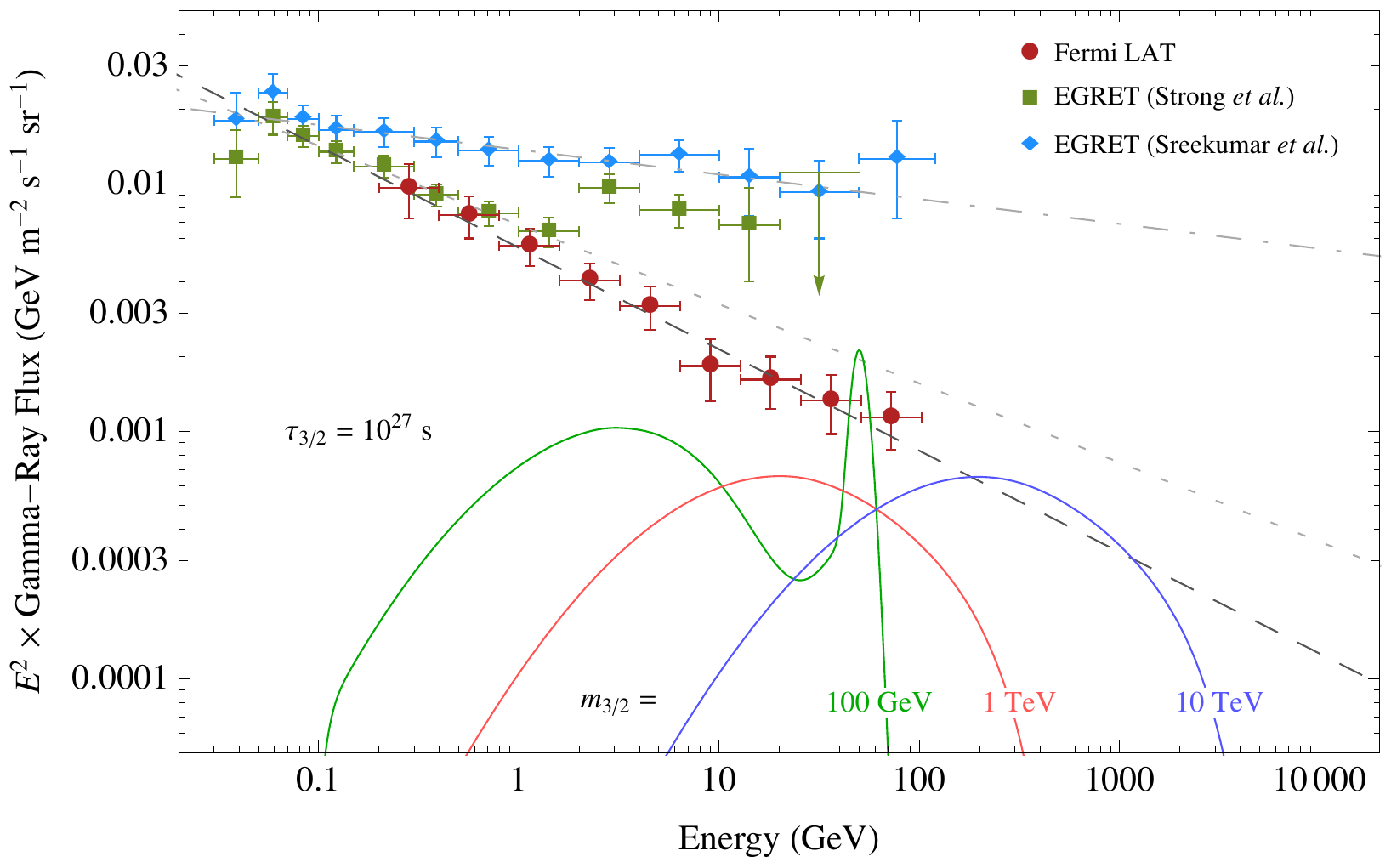}
  \caption{\label{antiprotongamma}Antiproton spectrum (\textit{left}) and gamma-ray spectrum (\textit{right}) expected from gravitino dark matter decays compared to data and the expected astrophysical background. From~\cite{DESY-THESIS-2011-039}.}
\end{figure}

In recent years several anomalies in the spectra of cosmic rays have been reported that cannot be explained by astrophysical secondary production and could be a hint for a signal from dark matter annihilation or decay. Indeed, the decay of gravitino dark matter leads to a hard spectrum of positrons (and electrons) that -- depending on the gravitino mass and lifetime -- could explain the rise in the positron fraction observed by the PAMELA experiment~\cite{arXiv:0810.4995}. This contribution could also be observed as a subdominant component in the absolute electron flux (see Figure~\ref{positronelectron}).
However, gravitino decays unavoidably produce associated fluxes of antiprotons and gamma rays that are only compatible with observations if the gravitino lifetime is larger than needed for an explanation of the PAMELA data (see Figure~\ref{antiprotongamma}). Therefore, a gravitino dark matter explanation of the PAMELA data is ruled out and additional astrophysical sources like pulsars are required to explain the observed positron fraction.

\begin{figure}[b]
\begin{minipage}[t]{0.49\textwidth}
  \includegraphics[width=\textwidth]{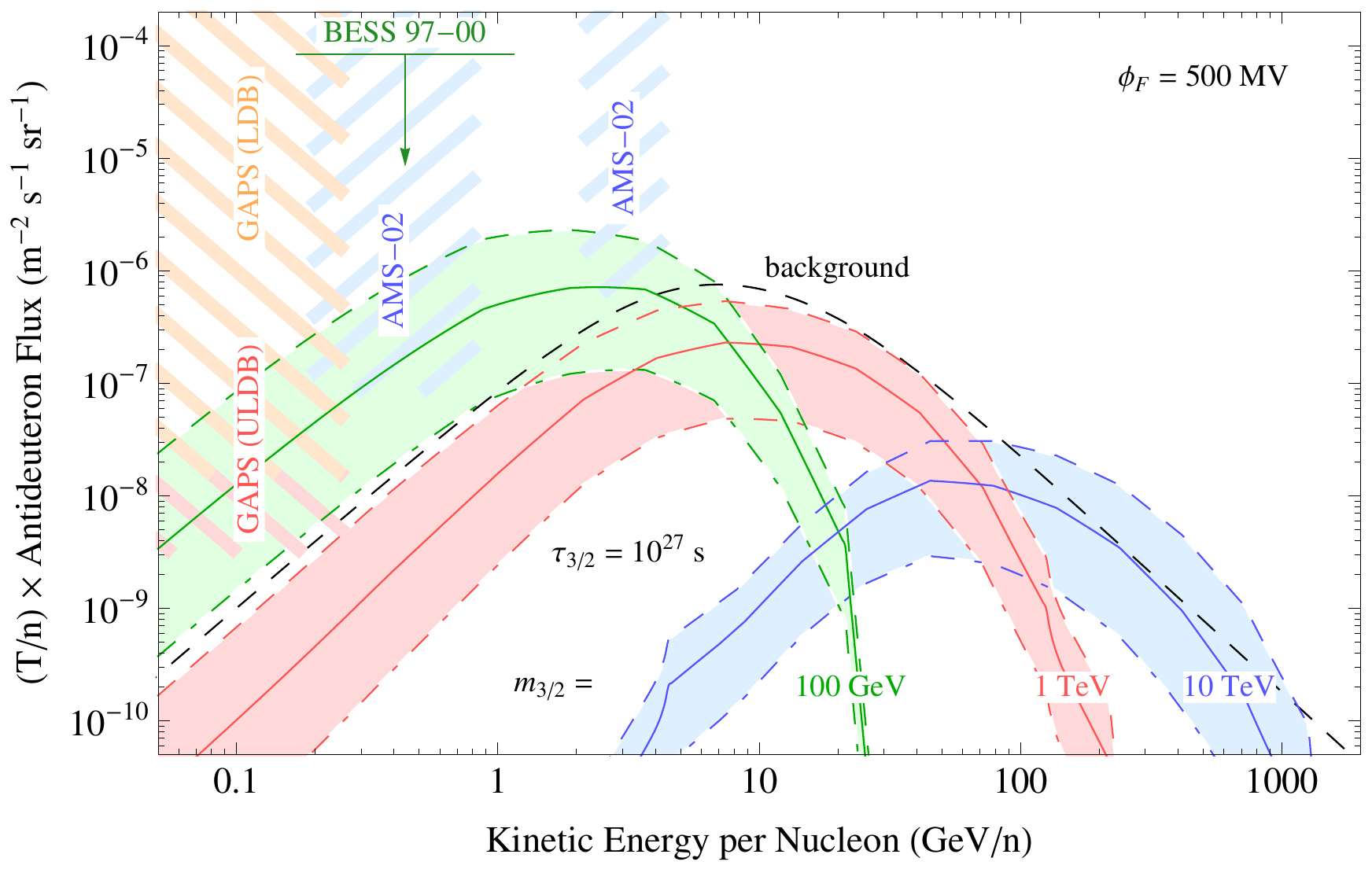}
  \caption{\label{antideuteron}Antideuteron spectrum expected from gravitino dark matter decays compared to sensitivities of forthcoming experiments and the expected astrophysical background.}
\end{minipage}
  \hfill
\begin{minipage}[t]{0.49\textwidth}
  \includegraphics[width=\textwidth]{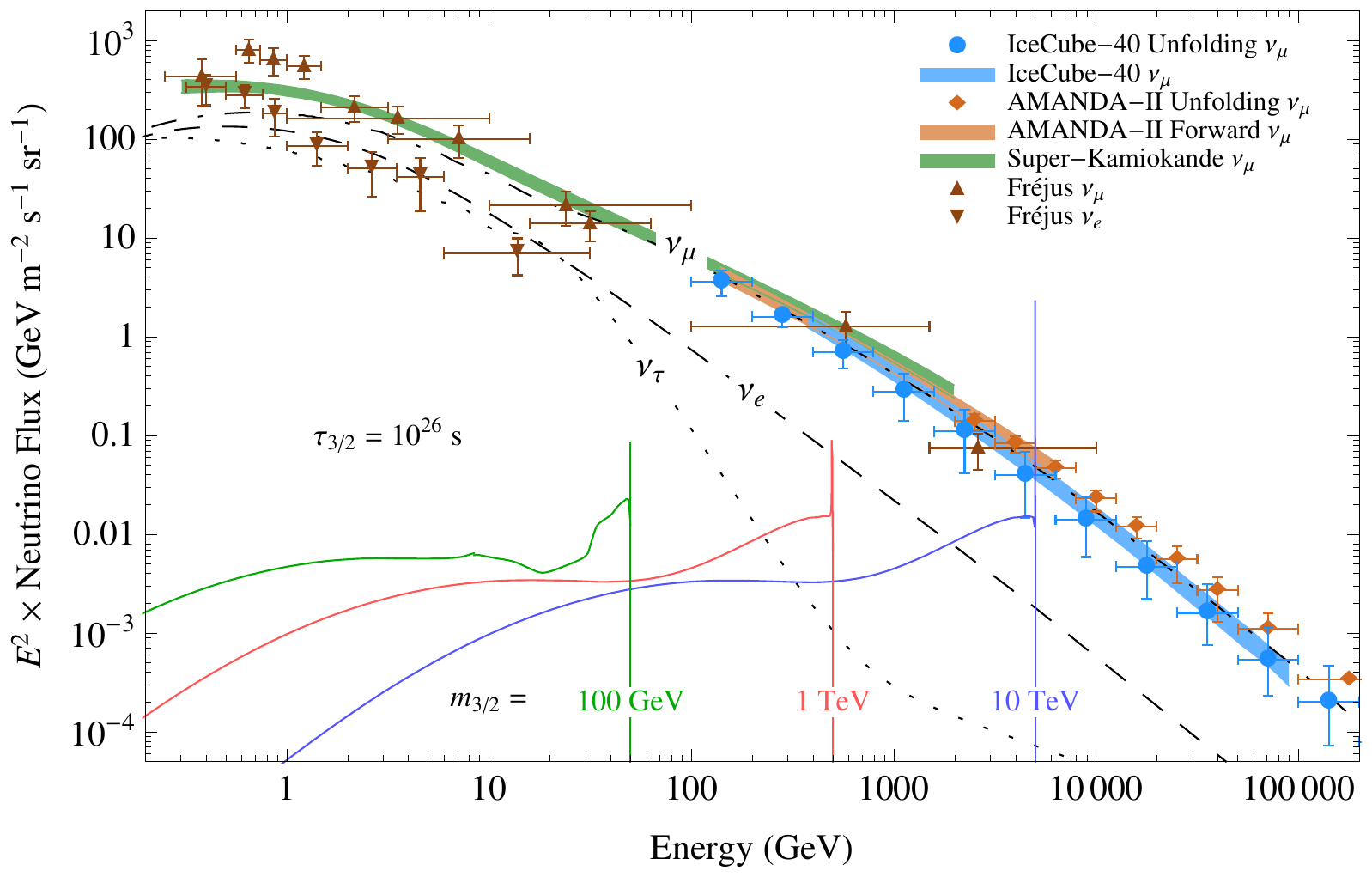}
  \caption{\label{neutrino}Neutrino spectrum expected from gravitino dark matter decays compared to data and the expected background of atmospheric neutrinos. Figure taken from~\cite{DESY-THESIS-2011-039}.}
\end{minipage}
\end{figure}

Also antideuterons can be formed in the fragmentation of gravitino decay products. The expected signal may exceed the astrophysical background especially for lower gravitino masses and can be tested with forthcoming experiments like AMS-02 and the GAPS balloon, thereby making antideuterons a valuable tool to probe light gravitinos (see Figure~\ref{antideuteron}).

\begin{figure}[b]
  \includegraphics[width=0.29\textwidth]{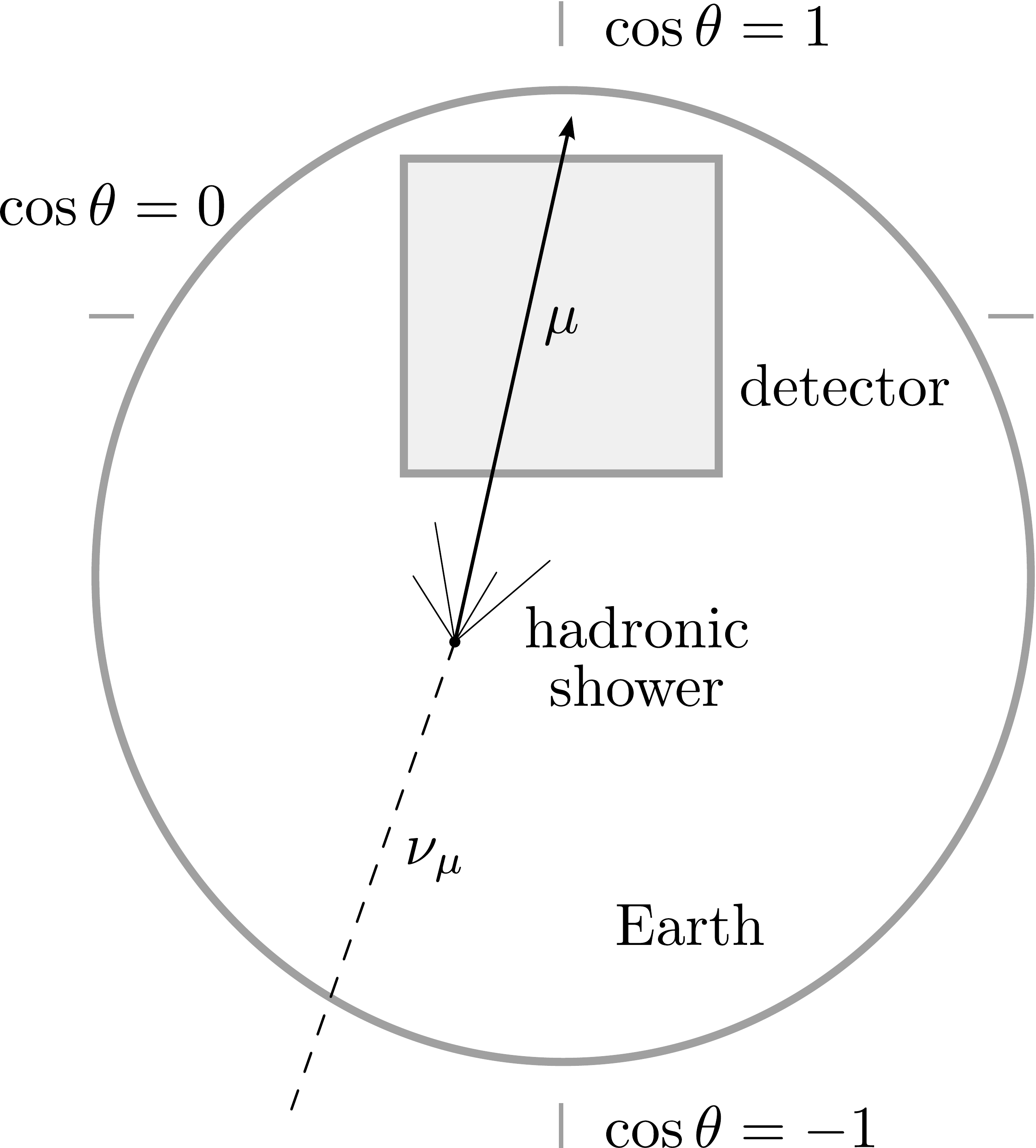}
  \hfill
  \includegraphics[width=0.34\textwidth]{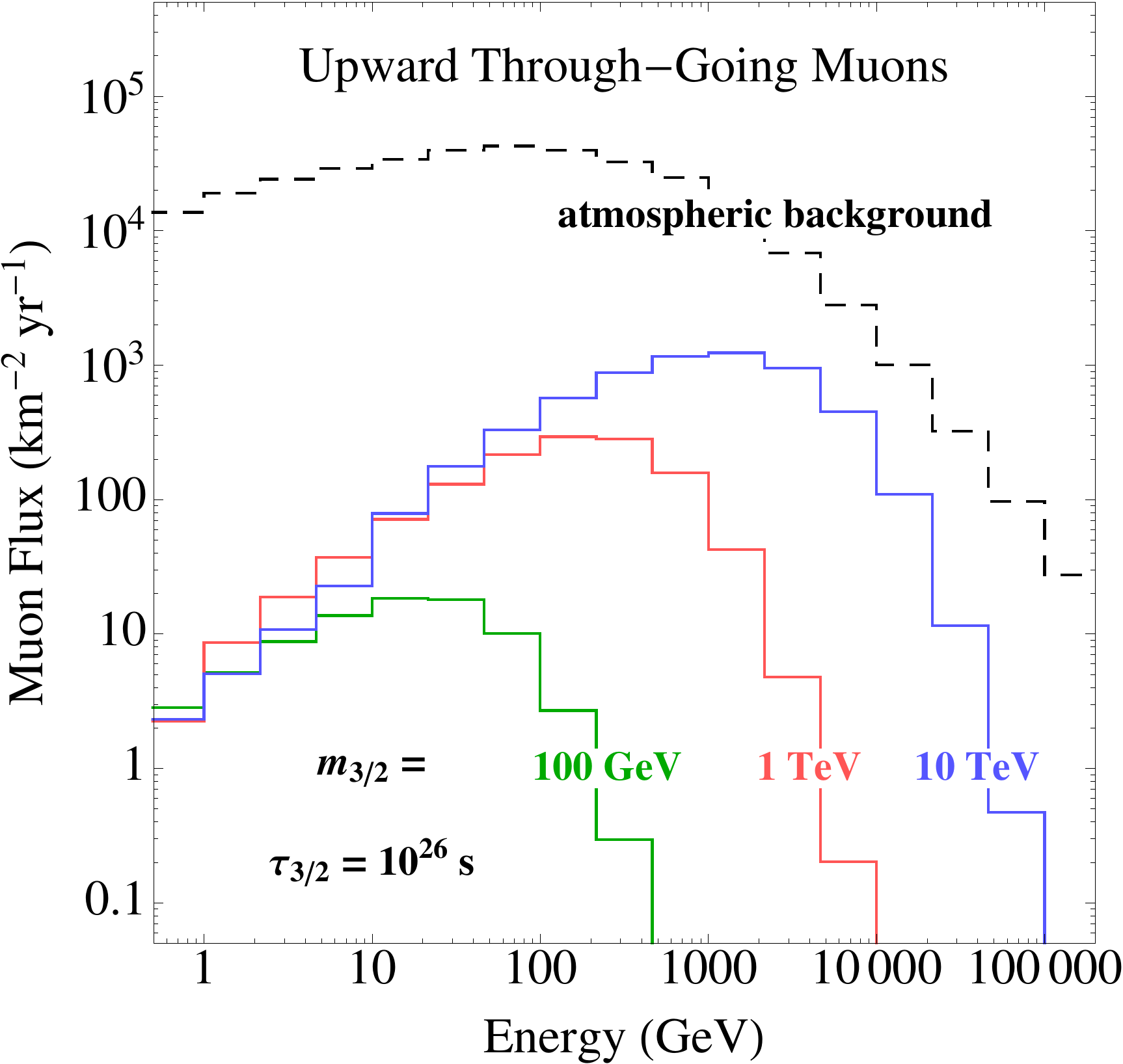}
  \hfill
  \includegraphics[width=0.34\textwidth]{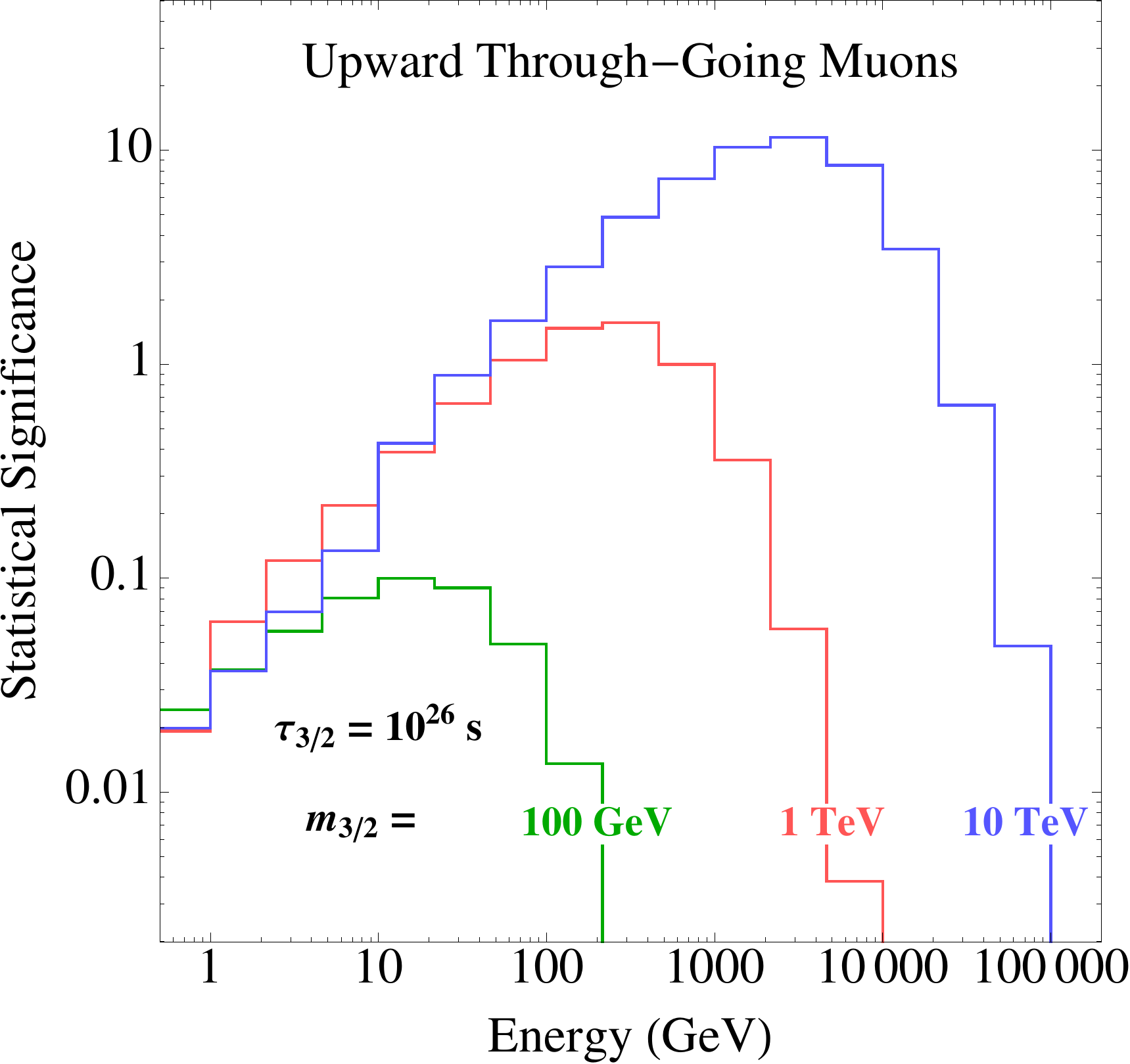}
  \caption{\label{muons}Event topology of neutrino-induced upward through-going muons (\textit{left}), muon spectrum with finite energy resolution (\textit{middle}) and statistical significance of the muon signal (\textit{right}). Figures taken from~\cite{DESY-THESIS-2011-039}.}
\end{figure}

\begin{figure}[t]
  \includegraphics[width=0.29\textwidth]{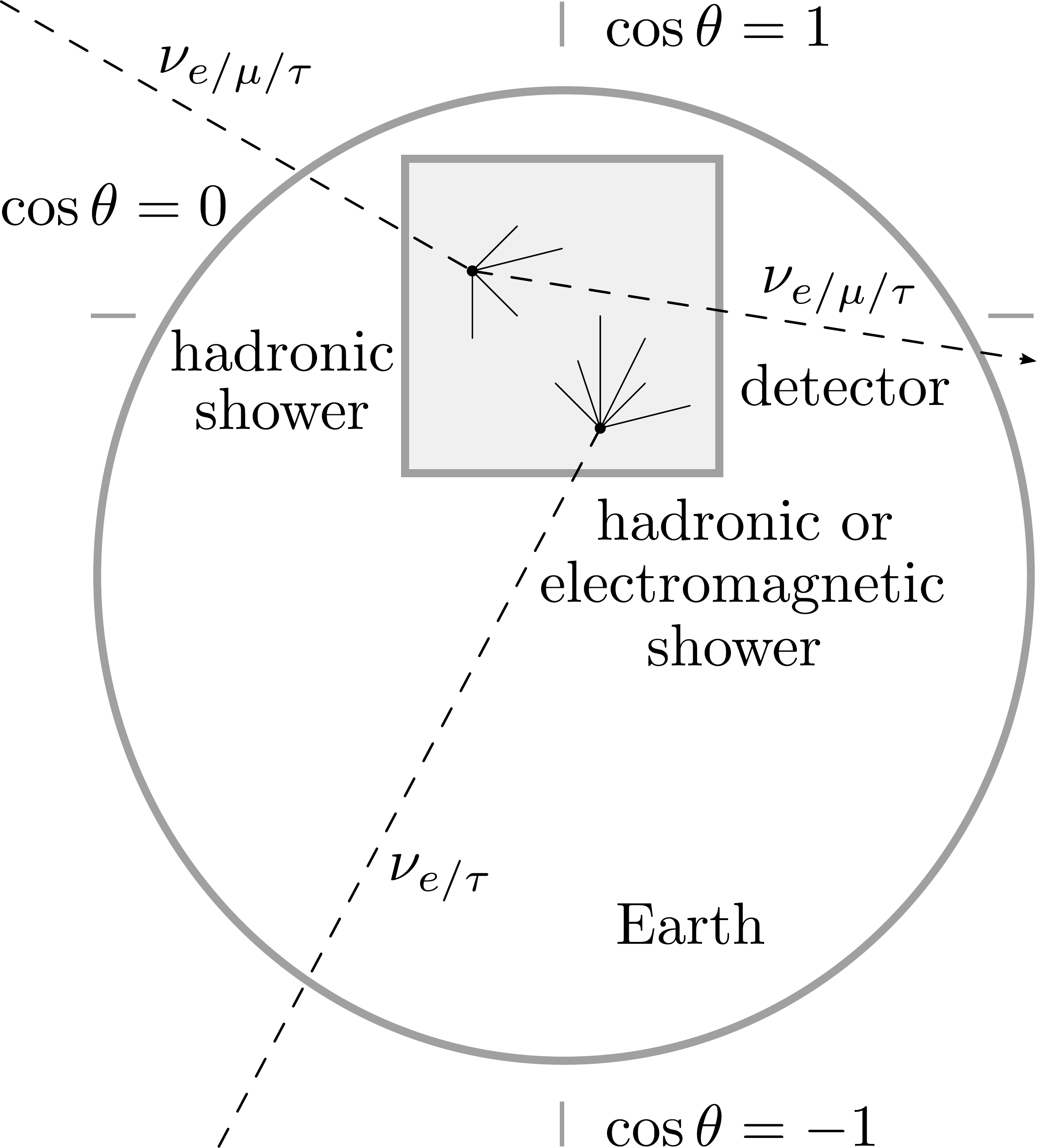}
  \hfill
  \includegraphics[width=0.34\textwidth]{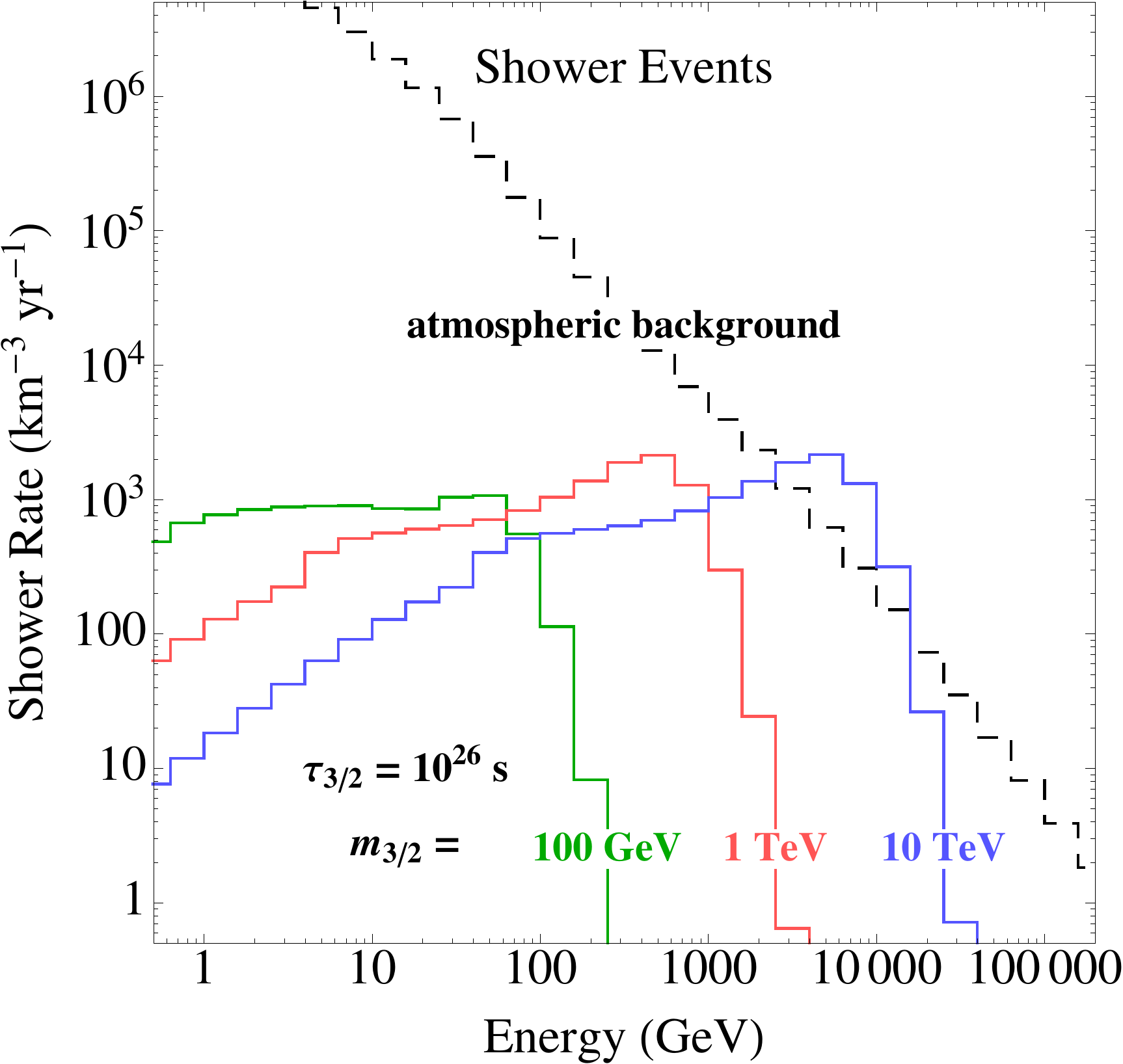}
  \hfill
  \includegraphics[width=0.34\textwidth]{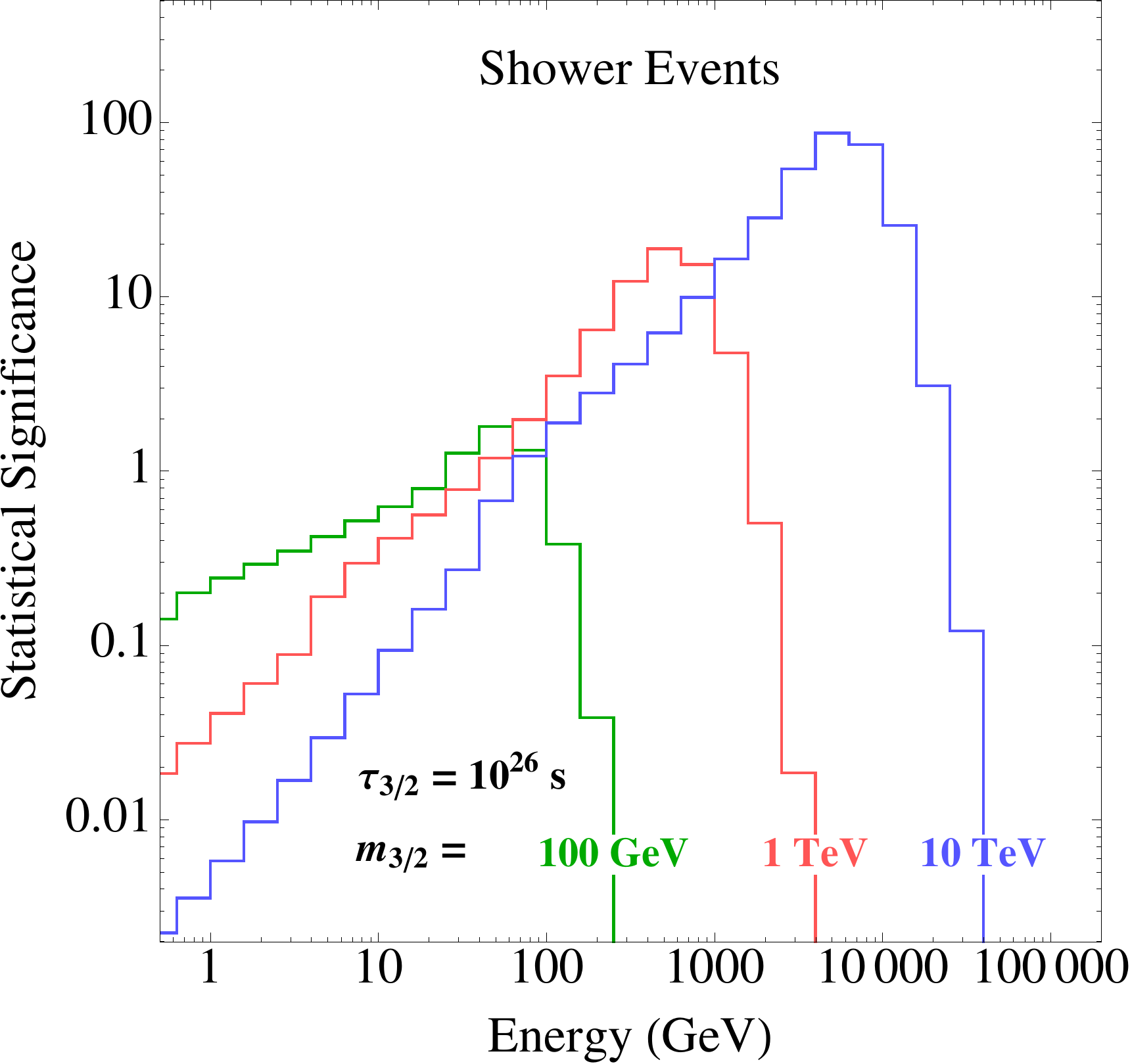}
  \caption{\label{showers}Event topology of neutrino-induced showers (\textit{left}), spectrum of showers with finite energy resolution (\textit{middle}) and statistical significance of the shower signal (\textit{right}). From~\cite{DESY-THESIS-2011-039}.}
\end{figure}

Another valuable channel are neutrinos that -- like gamma rays -- provide angular information that can be used to discriminate gravitino decay from dark matter annihilation~\cite{arXiv:0912.3521}. An intriguing feature of the expected signal is the line at the end of the spectrum that persists -- in contrast to gamma rays -- also for large gravitino masses (see Figure~\ref{neutrino}). In neutrino telescopes like IceCube neutrinos are mainly detected via the Cherenkov light of neutrino-induced upward through-going muons. Although the rather poor energy resolution of this channel obscures the expected spectral features, a statistical analysis might reveal a departure from the featureless spectrum of atmospheric neutrinos (see Figure~\ref{muons}). 
We find that a new detection channel, namely neutrino-induced showers, has two advantages compared to the muon signal: The dominant background of atmospheric muon neutrinos contributes less to this channel and its better energy resolution allows for a better discrimination of spectral features (see Figure~\ref{showers}). Although there are currently still problems to discriminate the shower signal from short muon tracks we conclude that this channel is the most promising for a detection of neutrinos from gravitino decays.

\begin{figure}[t]
  \includegraphics[width=0.37\textwidth]{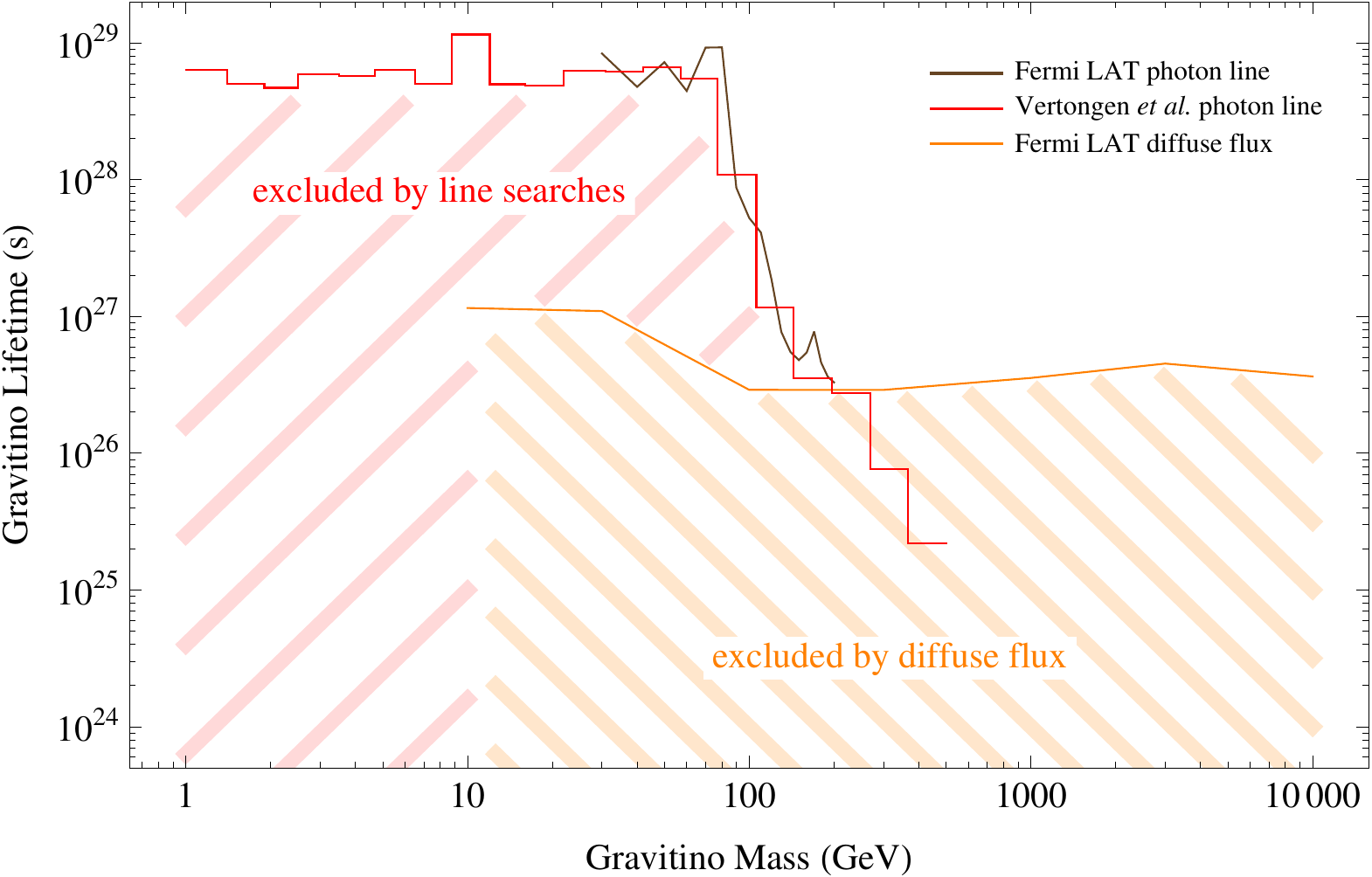}
  \hfill
  \includegraphics[width=0.37\textwidth]{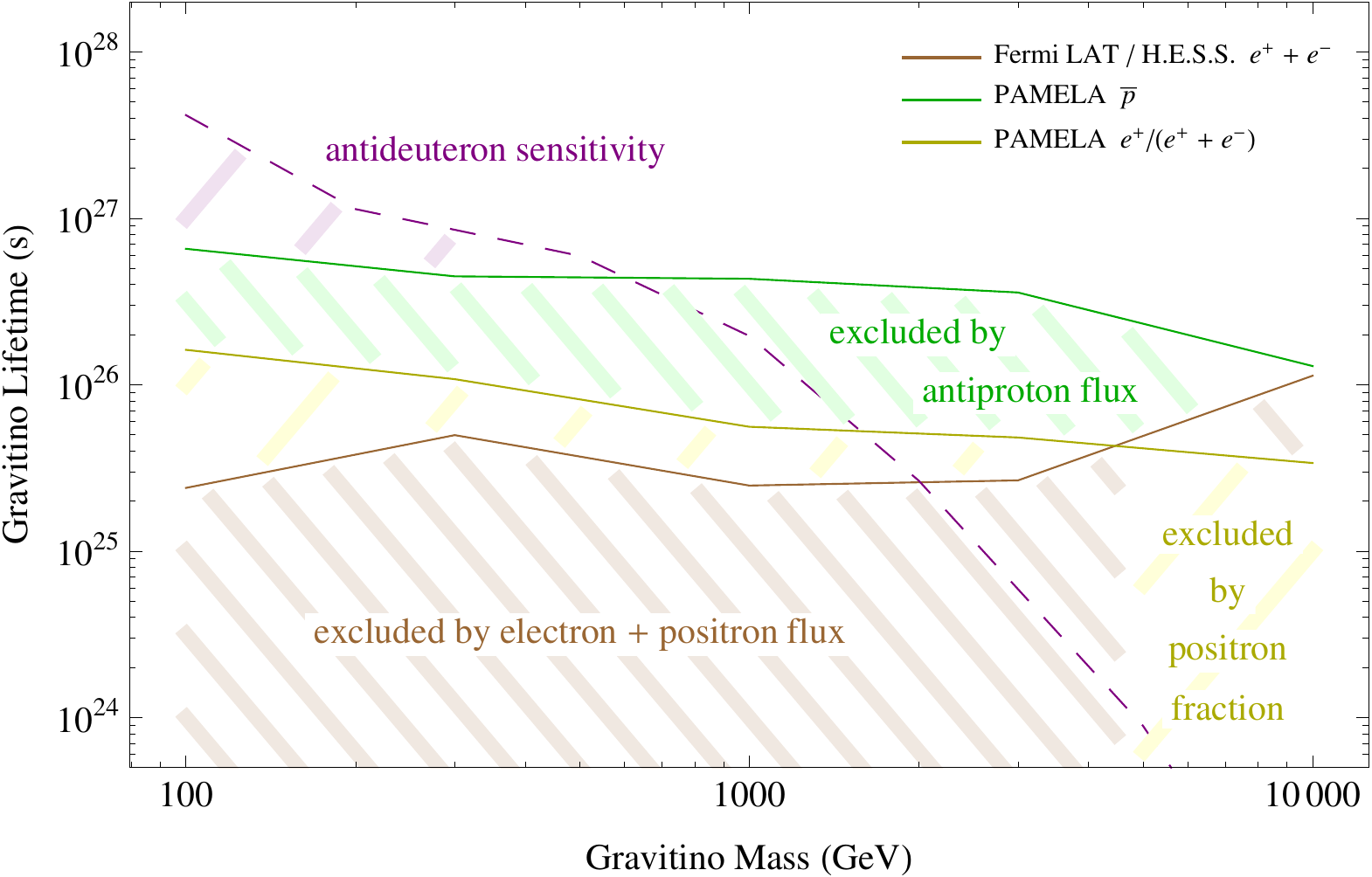}
  \hfill
  \includegraphics[width=0.244\textwidth]{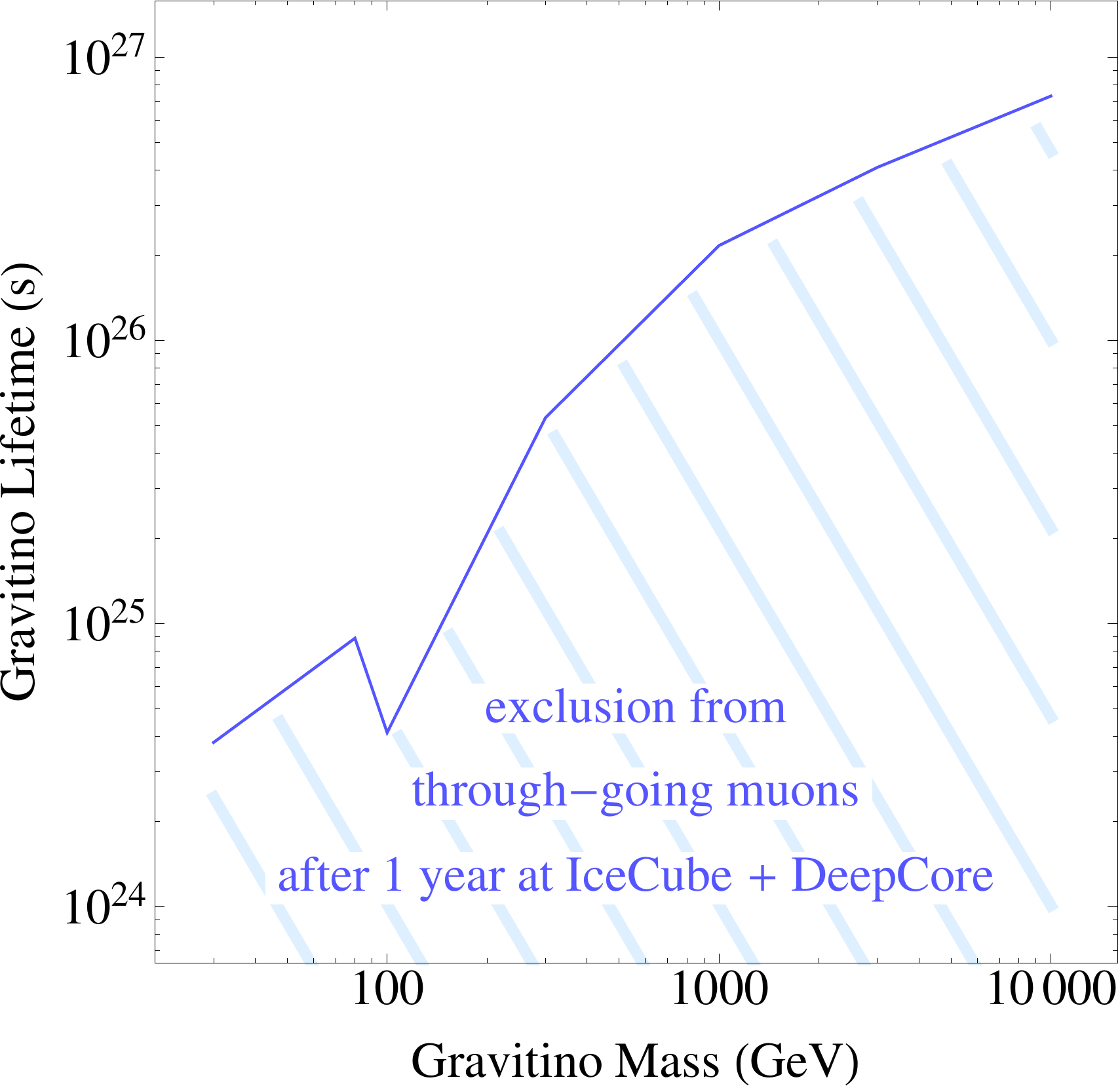}
  \caption{\label{lifetime}Estimated limits on the gravitino lifetime from gamma-ray observations (\textit{left}), observations of charged cosmic rays (\textit{middle}) and forthcoming observations of upward through-going muons at IceCube (\textit{right}). Figures taken from~\cite{DESY-THESIS-2011-039}.}
\end{figure}

The non-observation of gravitino decay signals in the spectra of various cosmic-ray species allows to estimate lower limits on the gravitino lifetime (see Figure~\ref{lifetime}). For light gravitinos the strongest limits come from gamma-ray line searches, while at larger masses the antiproton and diffuse gamma-ray limits are the most important. For very heavy gravitinos also neutrino bounds become significant. We expect that forthcoming antideuteron observations and searches based on neutrino-induced showers allow to improve these bounds in the future.

\section{Conclusions}

Unstable gravitino dark matter with a lifetime exceeding the age of the universe by many orders of magnitude is well motivated from cosmology. Although gravitino decays produce cosmic-ray spectra that could explain the PAMELA excess, this possibility is ruled out due to strong constraints on the associated production of antiprotons and gamma rays. Forthcoming antideuteron searches will be a valuable channel to search for signals of light gravitinos while neutrino telescopes like IceCube have the capability to probe heavy gravitinos -- in particular if new detection channels like shower events are employed. 

In conclusion, we find that the gravitino lifetime can be effectively constrained over a large range of gravitino masses using a multi-messenger approach. For an exhaustive discussion of the presented results see~\cite{DESY-THESIS-2011-039}.
\vspace{-3mm}

\ack

I am grateful to Laura Covi and Gilles Vertongen for useful discussions and collaboration on part of the presented work. I acknowledge the support of the DFG within the Collaborative Research Center 676, the Marie Curie ITN "UNILHC" under grant number PITN-GA-2009-237920 and the grant HEPHACOS S2009/ESP-1473 of the Comunidad de Madrid.

\end{document}